\documentclass[11pt]{article}

\pdfoutput=1

\usepackage[T1]{fontenc}
\usepackage[latin9]{inputenc}
\usepackage[a4paper]{geometry}
\usepackage[active]{srcltx}
\usepackage{amsmath}
\usepackage{amssymb}
\usepackage{esint}

\makeatletter

\usepackage{textcomp}

\pdfoutput=1 

\usepackage{jheppub}

\usepackage{etoolbox}
    
    \patchcmd{\maketitle}{\@fpheader}{}{}{}

\usepackage{amsfonts}

\usepackage{tikz}

\usepackage{bbm}

\renewcommand{\d}{\partial}

\renewcommand{\hat}{\widehat}

\newcommand{\be}{\begin{equation}}
\newcommand{\ee}{\end{equation}}
\newcommand{\bea}{\begin{eqnarray}}
\newcommand{\eea}{\end{eqnarray}}

\title{\boldmath The anisotropic chiral boson}

\author[a]{Oscar Fuentealba,}
\author[b]{Hern\'an A. Gonz\'alez,}
\author[c]{Miguel Pino,}
\author[a]{Ricardo Troncoso}

\affiliation[a]{Centro de Estudios Cient\'{i}ficos (CECs), Av. Arturo Prat 514, Valdivia, Chile}
\affiliation[b]{Facultad de Artes Liberales, Universidad Adolfo Ib\'a\~nez, Diagonal Las Torres 2640, Pe\~nalol\'en, Santiago, Chile}
\affiliation[c]{Departamento de F\'{i}sica, Universidad de Santiago de Chile, Avenida Ecuador 3493, Estaci\'on Central, 9170124, Santiago, Chile}

\emailAdd{fuentealba@cecs.cl}
\emailAdd{hernan.gonzalez@uai.cl}
\emailAdd{miguel.pino.r@usach.cl}
\emailAdd{troncoso@cecs.cl}

\preprint{CECS-PHY-19/03}

\abstract{
We construct the theory of a chiral boson with anisotropic scaling, characterized by a 
dynamical exponent $z$, whose action reduces to that of Floreanini and Jackiw in the isotropic
 case ($z=1$). The standard free boson with Lifshitz scaling is recovered when both 
chiralities are nonlocally combined. Its canonical structure and symmetries are also analyzed. 
As in the isotropic case, the theory is also endowed with a current algebra. 
Noteworthy, the standard conformal symmetry is shown to be still present, but realized in a
 nonlocal way. The exact form of the partition function at finite temperature is obtained from
 the path integral, as well as from the trace over $\hat{u}(1)$ descendants. It is essentially 
given by the generating function of the number of partitions of an integer into $z$-th powers,
 being a well-known object in number theory. Thus, the asymptotic growth of the number of states 
at fixed energy, including subleading corrections, can be obtained from the appropriate extension 
of the renowned result of Hardy and Ramanujan.}

\makeatother

\begin{document}
\maketitle \flushbottom

\newpage{}

\section{Introduction}

The self-dual free boson is a relativistic two-dimensional model describing
chiral massless excitations, that evolve according to the field equation
\begin{equation}
\dot{X}^{\pm}=\pm\partial_{x}X^{\pm}\,,
\end{equation}
where the sign corresponds to left or right movers. It enjoys a number
of interesting properties that appear in a variety of contexts, ranging
from string theory \cite{Gross:1984dd,Brink:1988nh} to the edge description
of the quantum Hall effect \cite{Stone:1992ft,Tong:2016kpv}. Finding
a suitable action principle for bosonic chiral fields requires to
deal with certain subtleties (see e.g., \cite{Siegel:1983es,Henneaux:1987hz,Henneaux:1988gg,Pasti:1996vs})
that in the case of a scalar were already addressed long ago by Floreanini
and Jackiw \cite{Floreanini:1987as}.

More recently, in the search for extensions of the holographic principle
(see e.g. \cite{Taylor:2008tg,Hartnoll:2009sz,Hartnoll:2011fn,Taylor:2015glc}),
there has been interest in analyzing quantum field theories that manifestly
break Lorentz invariance, but admit anisotropic scaling transformations
of the form
\begin{equation}
t\to\lambda^{z}t\,,\qquad x\to\lambda x\,,\label{eq:AnistropyZ}
\end{equation}
where $z$ is the dynamical exponent. The prototypical example of
such kind of field theories is the free boson with Lifshitz scaling
\cite{lifshitz1941theory}, which realizes the anisotropic symmetry
due to a quadratic term that contains higher spatial derivatives in
the action. Field theories enjoying these features are of interest
on their own (see e.g. \cite{Arav:2014goa,Chapman:2015wha,Arav:2016akx,Arav:2019tqm}),
as well as in different contexts related to condensed matter physics
\cite{Hertz:1976zz,Sachdev:1999zz,Bettelheim:2006un,Wiegmann:2011np,Sotiriadis:2017eot,Damia:2019bdx}.
Gravitational duals in which the anisotropic scaling symmetry is realized
through the (asymptotic) isometries of the so-called Lifshitz spacetimes
have been studied in e.g. \cite{Kachru:2008yh,Taylor:2008tg,Bertoldi:2009vn,Bertoldi:2009dt,Hartnoll:2009sz,DHoker:2010zpp,Hartnoll:2011fn,Gonzalez:2011nz,Hartnoll:2015faa,Taylor:2015glc,Matulich:2011ct,AyonBeato:2009nh}.
In the context of three--dimensional gravity, it has been shown that
the anisotropic scale invariance can also be induced by suitable choices
of boundary conditions \cite{Perez:2016vqo,Fuentealba:2017omf,Melnikov:2018fhb,Gonzalez:2018jgp,Grumiller:2019tyl,Ojeda:2019xih}.
In two spacetime dimensions, field theories with anisotropic scaling
have also been discussed in e.g. \cite{Chubukov:1994hp,PhysRevLett.93.066401,Cardy:1992tq,Hofman:2011zj,Melnikov:2018fhb}.

One of the main purposes of our work is the analysis of a new bosonic
quantum field theory in two dimensions that simultaneously combines
chirality with anisotropy. In the case of a single free boson, the
anisotropic scaling \eqref{eq:AnistropyZ} implies that the field
equation must be of the form
\begin{equation}
\dot{X}^{\pm}=\pm\sigma^{z-1}\partial_{x}^{z}X^{\pm}\,,\label{eq:Chiral-fe-Z}
\end{equation}
where $\sigma$ is a constant with units of length, so
that the speed of light becomes the unity for $z=1$. From \eqref{eq:Chiral-fe-Z}
it is clear that when $z$ takes even values, the modes can neither
suitably oscillate nor carry a single chirality (see below). Therefore,
hereafter the dynamical exponent $z$ is assumed to be given by an
odd integer.

The paper is organized as follows. In the next section we propose
an action principle for the anisotropic chiral boson that reduces
to the Floreanini-Jackiw action for $z=1$, and analyze some of its
properties. We also show how to combine both chiralities in order
to recover the standard free boson with Lifshitz scaling. In section
\ref{sec:sym}, we study the canonical structure and also explore
the global symmetries. It is found that, as in the isotropic case,
the theory is also endowed with a $\hat{u}(1)$ current algebra; and
remarkably, the conformal symmetry is still present, but turns out
to be non-local for $z>1$. In section \ref{sec:quantum}, the exact
form of the partition function at finite temperature is obtained first
from the path integral, and then from the trace over $\hat{u}(1)$
descendants. It is essentially given by the generating function of
the sequence of ``power partitions'', i.e., the number of partitions
of an integer into $z$-th powers. Therefore, the asymptotic growth
of the number of states at a given energy, including subleading corrections,
corresponds to power partitions associated to $\hat{u}(1)$ descendants,
and it can be obtained from the appropriate extension of the famous
result of Hardy and Ramanujan \cite{hardy1918asymptotic}.

\section{Action principle \label{Action principle}}

Let us consider the following action principle
\begin{equation}
S_{z}^{\pm}[X^{\pm}]=\int dt\,dx\left[\pm\partial_{x}X^{\pm}\partial_{t}X^{\pm}-\sigma^{(z-1)}\partial_{x}X^{\pm}\partial_{x}^{z}X^{\pm}\right],\label{acb}
\end{equation}
which is invariant under the anisotropic scaling in \eqref{eq:AnistropyZ},
provided that the field $X^{\pm}$ does not transform. Note that the
Floreanini-Jackiw theory is recovered for $z=1$.

It is worth highlighting that the action \eqref{acb} is invariant
under the gauge symmetry
\begin{equation}
X^{\pm}\rightarrow X^{\pm}+f^{\pm}(t)\,,\label{eq:Gauge symmetry}
\end{equation}
with arbitrary $f^{\pm}$ depending only on time. Indeed, the field
equation that follows from \eqref{acb} is
\begin{equation}
\partial_{x}\dot{X}^{\pm}=\pm\sigma^{z-1}\partial_{x}^{z+1}X^{\pm}\,,\label{eom1}
\end{equation}
and it can be readily integrated once, yielding 
\begin{equation}
\dot{X}^{\pm}=\pm\sigma^{z-1}\partial_{x}^{z}X^{\pm}+h^{\pm}(t)\,,\label{eom2}
\end{equation}
with $h^{\pm}$arbitrary. Thus, $h^{\pm}$ can always be gauged away
by virtue of the gauge transformation in \eqref{eq:Gauge symmetry}
with $h^{\pm}=\dot{f}^{\pm}$, so that \eqref{eom2} precisely reduces
to the field equation of the anisotropic chiral boson in \eqref{eq:Chiral-fe-Z},
i.e.,
\begin{equation}
\dot{X}^{\pm}=\pm\sigma^{z-1}\partial_{x}^{z}X^{\pm}\,.\label{eq:chiral-z}
\end{equation}
 The canonical analysis of the gauge symmetry \eqref{eq:Gauge symmetry}
is discussed in section \ref{gaugesym}.

According to \eqref{eq:chiral-z} the anisotropic chiral boson evolves
as a superposition of modes of the form
\begin{equation}
X_{k}^{\pm}=e^{i(kx\pm(i\sigma)^{z-1}k^{z}t)}\,,\label{modes}
\end{equation}
with a phase velocity given by $c_{p}^{\pm}=\pm(i\sigma k)^{z-1}$,
so that propagate in the same direction regardless the value of $k$.
As expected, different chiralities propagate in opposite directions.\footnote{Note that if $z$ were an even integer, the modes would not even propagate
since $c_{p}^{\pm}$ becomes purely imaginary, and also sensitive
to the sign of $k$. Besides, the second term in the action \eqref{acb}
turns out to be a boundary term in this case, so that by virtue of
the gauge symmetry, the field equation would actually be $\dot{X}^{\pm}=0$,
generically solved by $X^{\pm}=g^{\pm}(x)$. Thus, although an alternative
action principle could be written for even $z$, we do not consider
this possibility, since the free bosonic field has not chance of being
simultaneously propagating and chiral.} The magnitude of the phase velocity is then generically different
for each mode, reflecting the fact that the theory is not Lorentz
invariant (unless $z=1$). A wave packet then propagates in the same
direction than its composing modes, since the group velocity obeys
$c_{g}^{\pm}=zc_{p}^{\pm}$.

It is worth pointing out that a parity transformation $\mathcal{P}$,
defined by $x\rightarrow-x$, swaps the chirality. Indeed, the action
\eqref{acb} fulfills $\mathcal{P}[S_{z}^{\pm}]=-S_{z}^{\mp}$, regardless
the field $X^{\pm}$ transforms as a (pseudo-)scalar. Analogously,
chirality is also swapped by a time reversal transformation $\mathcal{T}$,
defined through $t\rightarrow-t$, i.e., $\mathcal{T}[S_{z}^{\pm}]=-S_{z}^{\mp}$.
Therefore, the joint action of parity and time reversal becomes a
symmetry of the anisotropic chiral boson action \eqref{acb} because
$\mathcal{P}\mathcal{T}[S_{z}^{\pm}]=S_{z}^{\pm}$.

\subsection{Recovering the standard free boson with Lifshitz scaling}

Here we show that two independent sectors with opposite chiralities
can be suitably combined in order to recover the standard free boson
with anisotropic scaling.

In two-spacetime dimensions, the action for a free boson with Lifshitz
scaling is given by \cite{lifshitz1941theory} (see also \cite{Melnikov:2018fhb})
\begin{equation}
I[\varphi]=\frac{1}{2}\int dt\,dx\left(\dot{\varphi}^{2}-\sigma^{2(z-1)}(\partial_{x}^{z}\varphi)^{2}\right)\,,\label{free}
\end{equation}
which is invariant under the anisotropic scaling \eqref{eq:AnistropyZ},
provided that the field transforms as
\begin{equation}
\varphi\rightarrow\lambda^{\frac{z-1}{2}}\varphi\,.\label{scaling}
\end{equation}
The field equations then correspond to an anisotropic version of the
wave equation 
\begin{equation}
\ddot{\varphi}-\sigma^{2(z-1)}\partial_{x}^{2z}\varphi=0\,.\label{wave}
\end{equation}
A key feature is that \eqref{wave} can be factorized as
\begin{equation}
D_{+}D_{-}\varphi=0\,,
\end{equation}
where $D_{\pm}\equiv\partial_{t}\pm\sigma^{(z-1)}\partial_{x}^{z}$
are linear differential operators, possessing well-defined scaling
behavior, since $D_{\pm}\rightarrow\lambda^{-z}D_{\pm}$. As a consequence
of the linearity of the equation --and up to zero modes-- the local
solution to \eqref{wave} is a d'Alembert-like superposition of two
non-interacting waves 
\begin{equation}
\varphi=\varphi^{+}+\varphi^{-}\,,\label{decomp}
\end{equation}
where 
\begin{equation}
D_{\mp}\varphi^{\pm}=0\,,
\end{equation}
which agree with the field equations of two anisotropic chiral bosons
$X^{\pm}$ with opposite chiralities in \eqref{eq:chiral-z}. However,
$\varphi^{\pm}$ inherit the scaling of $\varphi$ in \eqref{scaling},
while $X^{\pm}$ do not transform under the anisotropic scaling symmetry.

In order to see the precise link between the bosonic field $\varphi$
and the chiral fields $X^{\pm}$, it is useful to express the action
\eqref{free} in Hamiltonian form, which reads
\begin{equation}
I_{H}[\varphi,p]=\int dt\,dx\,\left[p\dot{\varphi}-\frac{1}{2}p^{2}-\frac{\sigma^{2(z-1)}}{2}(\d_{x}^{z}\varphi)^{2}\right]\,,\label{Ham-1}
\end{equation}
with $p$ the canonical momentum. Now, instead of using \eqref{decomp},
we focus on the following non-local field redefinition
\begin{equation}
\varphi=\frac{1}{\sqrt{\sigma^{(z-1)}}}\left(\partial_{x}^{\frac{1-z}{2}}X^{+}+\partial_{x}^{\frac{1-z}{2}}X^{-}\right)\,,\qquad p=\sqrt{\sigma^{(z-1)}}\left(\partial_{x}^{\frac{z+1}{2}}X^{+}-\partial_{x}^{\frac{z+1}{2}}X^{-}\right)\,,\label{redefphi}
\end{equation}
so that, up to zero modes, the dynamics of \eqref{Ham-1} can be equivalently
expressed in terms of the chiral fields $X^{\pm}$. This procedure
yields, up to boundary terms, the sum of two decoupled actions with
opposite chiralities
\begin{equation}
I_{H}[X^{+},X^{-}]=(-1)^{\frac{z-1}{2}}\left(S_{z}^{+}[X^{+}]+S_{z}^{-}[X^{-}]\right)\,,\label{eq:IH-sum}
\end{equation}
where $S_{z}^{\pm}[X^{\pm}]$ are given by \eqref{acb}.

Note that the anisotropic scaling transformation of the non-chiral
field $\varphi\rightarrow\lambda^{\frac{z-1}{2}}\varphi$ is recovered
from those of the chiral fields $X^{\pm}\rightarrow X^{\pm}$, by
virtue of the field redefinition \eqref{redefphi}.

It is also worth stressing that the field redefinition of $\varphi$
in \eqref{redefphi} is generically non-local, except for the case
$z=1$. Nonetheless, after carefully dealing with the boundary terms,
which can be consistently dropped out, the action in \eqref{eq:IH-sum}
becomes exclusively written in terms of local variables.

\section{Canonical structure and symmetries}

\label{sec:sym}

\subsection{Hamiltonian analysis}

\label{gaugesym}

Since the actions for chiral and antichiral fields were shown to be
connected through the parity operator $\mathcal{P}$ in section \ref{Action principle},
afterwards we drop the $\pm$ index. Thus, without loss of generality,
we continue with the analysis for a chiral field $X=X^{+}$. It is
also useful to assume that the chiral field is defined on a cylinder
of radius $l$, so that the spatial coordinate is rescaled as $x=l\phi$,
with $0\leq\phi<2\pi$.

In order to acquire a deeper understanding of the theory, including
its local and global symmetries, here we study its canonical structure.

The action of the anisotropic chiral boson in \eqref{acb} possesses
the following Hamiltonian form
\begin{equation}
I_{H}[\Pi,X,\lambda]=\int dtd\phi\,\left[\Pi\dot{X}-\alpha\partial_{\phi}X\partial_{\phi}^{z}X-\upsilon\left(\Pi-\partial_{\phi}X\right)\right]\,,\label{Hamiltonian_action}
\end{equation}
where $\Pi$ stands for the canonical momentum, and $\upsilon$ is
a Lagrange multiplier. The Hamiltonian equations of motion then read
\begin{equation}
\dot{X}=\upsilon\,,\quad\dot{\Pi}=2\alpha\partial_{\phi}^{z+1}X-\partial_{\phi}\upsilon\,,\quad\Pi-\partial_{\phi}X=0\,,\label{Hamiltonian_eom}
\end{equation}
being equivalent to \eqref{eom1}. The latter equality corresponds
to a primary constraint that arises from the definition of the momentum,
that is included in the Hamiltonian action \eqref{Hamiltonian_action}
along with the Lagrange multiplier. The canonical Poisson bracket
can then be read off from the kinetic term of the action
\begin{equation}
\{X(\phi),\Pi(\phi')\}=\delta(\phi-\phi')\,.\label{canonical_poisson}
\end{equation}
In terms of Fourier modes, the fields expand as
\begin{equation}
X(t,\phi)=\sum_{n=-\infty}^{\infty}e^{in\phi}X_{n}(t)\,,\quad\Pi(t,\phi)=\sum_{n=-\infty}^{\infty}e^{in\phi}\Pi_{n}(t)\,,\quad\upsilon=\sum_{n=-\infty}^{\infty}e^{in\phi}\upsilon_{n}(t)\,,
\end{equation}
so that the Poisson bracket \eqref{canonical_poisson} reads
\begin{equation}
\{X_{n},\Pi_{m}\}=\frac{1}{2\pi}\delta_{n,-m}\,.\label{canonical_poisson_n}
\end{equation}
Analogously, the smeared constraint $\theta[\upsilon]\equiv\int d\phi\,\upsilon\left(\Pi-\partial_{\phi}X\right)$,
can be written as
\begin{equation}
\theta[\upsilon]=\sum_{n=-\infty}^{\infty}\upsilon_{n}\theta_{n}\,,
\end{equation}
with
\begin{equation}
\theta_{n}=2\pi\left(\Pi_{-n}+inX_{-n}\right)\,.
\end{equation}
It is then straightforward to check that
\begin{equation}
\{\theta_{n},\theta_{m}\}=4\pi in\delta_{n+m,0}\,,\label{lolo}
\end{equation}
implying that $\theta_{0}=2\pi\Pi_{0}$ is a first class constraint,
while $\theta_{n}$ for $n\neq0$ are of second class.

Note that the time evolution of the constraints reads
\begin{equation}
\dot{\theta}_{n}=4\pi in\left[\alpha(in)^{z}X_{-n}-\upsilon_{-n}\right]\,.
\end{equation}
Thus, for the mode $n=0$ the consistency of the constraint does not
add any new condition, while for $n\neq0$ it fixes the Lagrange multiplier
$\upsilon_{-n}$. Consequently, the first class constraint $\theta[\upsilon_{0}]$
generates gauge transformations, given by
\begin{equation}
\delta X=\{X,\theta[\upsilon_{0}]\}=\upsilon_{0}(t)\,,\label{eq:Gauge-symm}
\end{equation}
corresponding to the gauge symmetry in \eqref{eq:Gauge symmetry},
with $\upsilon_{0}(t)=f^{+}(t)$.

In order to deal with second class constraints, we use the Dirac bracket,
defined by
\begin{equation}
\{X_{n},\Pi_{m}\}_{D}=\{X_{n},\Pi_{m}\}-\sum_{k\neq0\,l\neq0}\{X_{n},\theta_{k}\}C_{kl}\{\theta_{l},\Pi_{m}\}\,,
\end{equation}
where $C_{kl}$ is the inverse of the matrix $\{\theta_{k},\theta_{l}\}$
in \eqref{lolo}, given by 
\begin{equation}
C_{kl}=\frac{1}{4\pi il}\delta_{k,-l}\,.
\end{equation}
In particular, it is simple to verify that the Dirac bracket of the
dynamical fields $X_{n}$ and $\Pi_{m}$ (with $n,m\neq0$) reads
\begin{equation}
\{X_{n},\Pi_{m}\}_{D}=\frac{1}{4\pi}\delta_{n,-m}\,,
\end{equation}
being equivalent to
\begin{equation}
\{X(\phi),\Pi(\phi')\}_{D}=\frac{1}{2}\delta(\phi-\phi')\,,\label{eq:Dirac-bracket}
\end{equation}
excluding zero modes. Therefore, the second class constraints can
be strongly imposed in a consistent way, so that
\begin{equation}
\{X(\phi),\partial_{\phi'}X(\phi')\}_{D}=\frac{1}{2}\delta(\phi-\phi')\,.\label{eq:Symplectic-bracket}
\end{equation}

It is worth pointing out that the Dirac bracket in \eqref{eq:Symplectic-bracket}
precisely agrees with the derivative of the inverse of the symplectic
form that can be obtained directly from \eqref{acb} (see e.g. \cite{Faddeev:1988qp}).
The latter bracket also helps in order to directly perform the analysis
of the global symmetries and their corresponding algebras, that is
carried out next.

\subsection{Global symmetries \label{gsym}}

\subsubsection{Kinematical symmetries \& Lifshitz algebra in $2d$}

Apart from the scaling symmetry in \eqref{eq:AnistropyZ}, with $X\rightarrow X$,
the action of the anisotropic chiral boson \eqref{acb} is also invariant
under displacements in space and time. This set of kinematical symmetries
is spanned by the vector field
\begin{equation}
Y=(a^{t}+bzt)\partial_{t}+(a^{\phi}+b\phi)\partial_{\phi}\,,\label{eq:KIlling-anisotropic}
\end{equation}
where $a^{t}$, $a^{\phi}$ and $b=\log\lambda$ are independent constants.
The infinitesimal transformation of the field $X$ is then given by
\begin{equation}
\delta X=Y^{\mu}\partial_{\mu}X\,,\label{kine}
\end{equation}
so that the associated Noether charges can be readily found to be
\begin{equation}
Q[a^{t},a^{\phi},b]=\int d\phi\left[(a^{t}+bzt)\alpha\partial_{\phi}X\partial_{\phi}^{z}X+(a^{\phi}+b\phi)(\partial_{\phi}X)^{2}\right]\,.
\end{equation}
The energy then agrees with the Hamiltonian in \eqref{Hamiltonian_action},
i.e.,
\begin{align}
H=Q[1,0,0] & =\alpha\int d\phi\,\d_{\phi}X\d_{\phi}^{z}X\,,\label{Ham}
\end{align}
while the momentum $P$ and the anisotropic scaling generator $D$
are identified as
\begin{align}
P=Q[0,1,0] & =\int d\phi\,\left(\partial_{\phi}X\right)^{2}\,,\label{eq:Momentum}\\
D=Q[0,0,1] & =\int d\phi\left[zt\alpha\partial_{\phi}X\partial_{\phi}^{z}X+\phi(\partial_{\phi}X)^{2}\right]\,.
\end{align}

The generators of the kinematical symmetries can then be shown to
span the Lifshitz algebra in $1+1$ dimensions, i.e.,
\begin{equation}
\{H,P\}_{D}=0\,,\qquad\{P,D\}_{D}=P\,,\qquad\{H,D\}_{D}=zH\,,
\end{equation}
in agreement with the Lie bracket of the Killing vectors \eqref{eq:KIlling-anisotropic}.

\subsubsection{$\hat{u}(1)$ current algebra}

Since the field equation of the anisotropic chiral boson is linear,
the superposition principle certainly holds. This simple fact can
be realized as a Noetherian symmetry of the action \eqref{acb}, whose
associated conserved charges satisfy the $\hat{u}(1)$ current algebra.
Those are ``shift symmetries'' given by
\begin{equation}
\delta X=\eta\,,\label{deltaX}
\end{equation}
where $\eta$ is regarded as a parameter that fulfills the field equation.
Assuming periodic boundary conditions, $\eta$ is given by a superposition
of modes of the form
\begin{equation}
\eta_{n}=e^{i(n\phi+(-1)^{\frac{z-1}{2}}\alpha n^{z}t)}\,.\label{eq:eta_n}
\end{equation}
The corresponding conserved charges that span the shift symmetries
\eqref{deltaX} then read
\begin{equation}
K[\eta]=2\int\eta\partial_{\phi}X\,d\phi\,,\label{K}
\end{equation}
and it is then simple to verify that their modes $K_{n}\equiv\frac{1}{2}K[\eta_{n}]$
fulfill the $\hat{u}(1)$ current algebra 
\begin{equation}
i\{K_{n},K_{m}\}_{D}=\pi n\delta_{n+m,0}\,.\label{U1}
\end{equation}
It is worth highlighting that, although the modes $K_{n}$ depend
on the dynamical exponent $z$, the $\hat{u}(1)$ current algebra
does not; and hence the algebra coincides with that of the standard
(isotropic) chiral boson for $z=1$. These generators play a leading
role in the construction of the Hilbert space for the quantum theory
(see section \ref{sec:quantum}).

\subsubsection{Conformal algebra from a nonlocal symmetry}

Following the common lore (see e.g. \cite{DiFrancesco:1997nk}) the
generators of the standard conformal symmetry can be constructed out
from those of the $\hat{u}(1)$ currents by means of the Sugawara
construction
\begin{equation}
L_{n}=\frac{1}{2\pi}\sum_{j=-\infty}^{\infty}K_{j}K_{n-j}\,,\label{virasoromode}
\end{equation}
so that their Dirac brackets span a single chiral copy of the Witt
algebra
\begin{equation}
i\{L_{n},L_{m}\}_{D}=(n-m)L_{n+m}\,.
\end{equation}
In the anisotropic case ($z>1$), it is amusing to verify that the
conformal algebra, spanned by $L_{n}$, is still present, but realized
as a nonlocal symmetry of the action \eqref{acb}. Indeed, the corresponding
transformation law of the field is given by
\begin{equation}
\delta X(\phi)=\int d\phi'f(\phi,\phi',t)\partial_{\phi'}X(\phi')\,,\label{deltavirasoro}
\end{equation}
where $f(\phi,\phi',t)=f(\phi',\phi,t)$ can be expressed in terms
of the modes $\eta_{j}(\phi,t)$ in \eqref{eq:eta_n}, so that
\begin{equation}
f(\phi,\phi',t)=\frac{1}{2\pi}\sum_{n,j=-\infty}^{\infty}\epsilon_{n}\eta_{j}(\phi,t)\eta_{n-j}(\phi',t)\,.\label{eq:funcion-f}
\end{equation}
Therefore, the conserved quantity associated to the nonlocal conformal
symmetry \eqref{deltavirasoro} is given by
\begin{align}
L[\epsilon_{n}] & =\int d\phi\,d\phi'f\partial_{\phi}X(\phi)\partial_{\phi'}X(\phi')\,,\label{cool}\\
 & =\sum_{n=-\infty}^{\infty}\epsilon_{n}L_{n}\,.
\end{align}
It is worth highlighting that the zero mode actually corresponds to
a local symmetry, since $L_{0}=P$, where $P$ stands for the momentum
generator in \eqref{eq:Momentum} (instead of the energy $H$). Indeed,
for the zero mode, the function in \eqref{eq:funcion-f} reduces to
\begin{equation}
f(\phi,\phi',t)=\epsilon_{0}{\displaystyle \sum_{j=-\infty}^{\infty}}\eta_{j}(\phi,t)\eta_{-j}(\phi',t)=2\pi\epsilon_{0}\delta(\phi-\phi')\,,
\end{equation}
and hence $L[\epsilon_{0}]=2\pi\epsilon_{0}\int d\phi\,\left(\partial_{\phi}X\right)^{2}$.
For $z>1$, the remaining modes necessarily span a nonlocal symmetry.

As expected, in the isotropic case ($z=1$) the conformal transformation
\eqref{deltavirasoro} becomes local. In fact, in this case the function
in \eqref{eq:funcion-f} reads
\begin{equation}
f(\phi,\phi',t)=\sum_{n=-\infty}^{\infty}\epsilon_{n}e^{in(\phi'+\alpha t)}\delta(\phi-\phi')=\epsilon(\phi'+\alpha t)\delta(\phi-\phi')\,,
\end{equation}
and hence \eqref{deltavirasoro} and \eqref{cool} reduce to the transformation
law and the generators of the conformal symmetry of the Floreanini-Jackiw
action \cite{Floreanini:1987as}, given by
\begin{equation}
\delta X(\phi)=\epsilon(\phi+\alpha t)\partial_{\phi}X(\phi)\,,
\end{equation}
\begin{equation}
L[\epsilon]=\int d\phi\,\epsilon(\phi+\alpha t)(\partial_{\phi}X(\phi))^{2}\,,
\end{equation}
respectively.

\section{Quantum aspects}

\label{sec:quantum}

In order to make the passage to the quantum theory, here we represent
the vacuum-to-vacuum transition amplitude in terms of a path integral
in the Hamiltonian formulation. This approach is particularly well
suited to deal with constrained systems, as it is the case of the
theory under discussion. Among its advantages, one has a more precise
control of the integration measure. We begin working in Lorentzian
signature and then perform a Wick rotation in order to obtain the
partition function at finite temperature.

In presence of first class constraints $\psi_{a}\approx0$ with gauge
fixing conditions $\chi_{a}\approx0$, together with second class
constraints $\theta_{n}\approx0$, the vacuum-to-vacuum transition
amplitude $W$ for a Hamiltonian system is given by \cite{Senjanovic:1976br,Faddeev:1980be,Henneaux:1992ig}
\begin{equation}
W=\int\mathcal{D}X\mathcal{\mathcal{D}}\Pi\exp\left[i\int dt\left(\Pi_{i}\dot{X}_{i}-H\right)\right]\,,\label{eq:W}
\end{equation}
where the integration measure reads
\begin{equation}
\mathcal{D}X\mathcal{D}\Pi=\prod_{a}\delta(\chi_{a})\delta(\psi_{a})\det\{\chi_{a},\psi_{b}\}\prod_{k}\delta(\theta_{k})\left(\det\{\theta_{n},\theta_{m}\}\right)^{1/2}\prod_{i}\mathcal{D}X_{i}\,\mathcal{D}\Pi_{i}\,.\label{mea}
\end{equation}
In order to adapt \eqref{eq:W} and \eqref{mea} to our case, the
first step is decomposing the Hamiltonian action in \eqref{Hamiltonian_action}
(without the constraints) in terms of Fourier modes,
\begin{equation}
I_{H}[\Pi,X,0]=2\pi\int dt\,\Pi_{0}\dot{X}_{0}+2\pi\sum_{n\neq0}\int dt\left(\Pi_{n}\dot{X}_{-n}-\frac{1}{l}n^{z+1}X_{n}X_{-n}\right)\,,\label{eq:Hamiltonian action Z}
\end{equation}
where the zero mode has been manifestly isolated from the remaining
modes\textbf{ }and we have made the choice $\alpha=(-1)^{(z-1)/2}l^{-1}$.
As explained in section \ref{gaugesym}, there is a single first class
constraint given by $\psi=2\pi\Pi_{0}$, which can be gauge fixed
according to $\chi=X_{0}$. Thus, replacing these conditions on the
first part of \eqref{mea} and integrating over $\Pi_{0}$ and $X_{0}$,
we obtain that that the zero mode sector completely disappears from
the amplitude. Furthermore, integrating over $\Pi_{n}$ yields

\begin{equation}
W=\int\prod_{n\neq0}\mathcal{D}X_{n}(\det\{\theta_{k},\theta_{l}\})^{1/2}e^{iS[X]}\,,\label{lklk}
\end{equation}
where $S[X]$ corresponds to strongly imposing the second class constraints
$\Pi_{n}=inX_{n}$ in what remains of the action \eqref{eq:Hamiltonian action Z},
i.e.,
\begin{equation}
S[X]=2\pi\sum_{n\neq0}\int dt\left(inX_{n}\dot{X}_{-n}-\frac{1}{l}n^{z+1}X_{n}X_{-n}\right)\,.\label{eq:SX-Lor}
\end{equation}

Note that the integration measure in \eqref{lklk} manifestly carries
a factor $\left(\det\{\theta_{k},\theta_{l}\}\right)^{1/2}$, due
to the presence of second class constraints, which can be readily
evaluated by virtue of \eqref{lolo} (see below).

\subsection{Partition function}

At finite temperature $T=\beta^{-1}$, the partition function $Z[\beta]$
can be obtained performing a Wick rotation of $W$, with $t=-iy$
and summing over periodic configurations with period $\beta$, i.e.,\footnote{For simplicity, we do not consider winding modes on both inequivalent
cycles of the torus.}
\begin{equation}
X_{n}(y)=\sum_{m}X_{n,m}e^{\frac{2\pi imy}{\beta}}\,.
\end{equation}
In terms of these variables, the Euclidean continuation of the action
\eqref{eq:SX-Lor} reads

\begin{equation}
iS_{E}[X]=-\frac{1}{2}\sum_{n\neq0,m,k,s}(-4\pi i n\delta_{n+m,0})\left(2\pi s+i\frac{\beta}{l}n^{z}\right)\delta_{k+s,0}X_{m,k}X_{n,s}\,.\label{eq:SX-Euclidean}
\end{equation}
Performing the remaining Gaussian integrals, it can be explicitly
seen that the contribution coming from the term $4\pi i n\delta_{m+n,0}$
in the Euclidean action \eqref{eq:SX-Euclidean}, and the factor $(\det\{\theta_{k},\theta_{l}\})^{1/2}$
that comes from the integration measure in \eqref{lklk}, precisely
cancel out. Thus, in terms of the modular parameter of the torus $\tau=i\frac{\beta}{2\pi l}$,
the partition function is given by
\begin{equation}
Z[\tau]=\mathcal{N}\prod_{n\neq0}\prod_{s=-\infty}^{\infty}(s+n^{z}\tau)^{-1/2}\,,\label{eq3}
\end{equation}
where the normalization factor $\mathcal{N}$ does not depend on the temperature.

In order to deal with the divergent infinite product in \eqref{eq3},
it is useful to consider the average energy of the system (see e.g.,
\cite{Cotler:2018zff}), given by
\begin{equation}
\langle E\rangle=-\frac{i}{2\pi l}\frac{\partial}{\partial\tau}\log Z=\frac{i}{2l}\sum_{n=1}^{\infty}n^{z}\cot(\pi n^{z}\tau)\,,
\end{equation}
which can be evaluated by virtue of the $\zeta$-function regularization.
Indeed,
\begin{align}
\frac{\partial}{\partial\tau}\log Z & =-\pi\sum_{n=1}^{\infty}n^{z}[\cot(\pi n^{z}\tau)+i]+i\pi\sum_{n=1}^{\infty}n^{z}\,,\\
 & =\frac{\partial}{\partial\tau}\left[\sum_{n=1}^{\infty}\log\left(\frac{1}{1-q^{n^{z}}}\right)\right]+i\pi\zeta(-z)\,,
\end{align}
where $q=\exp(2\pi i\tau)$, and $\zeta(s)={\displaystyle \sum_{n=1}^{\infty}}n^{-s}$
is the Riemann zeta function.

Therefore, the partition function is given by
\begin{equation}
Z[\tau]=q^{\frac{1}{2}\zeta(-z)}\prod_{n=1}^{\infty}\frac{1}{1-q^{n^{z}}}\,,\label{chz}
\end{equation}
where $E_{0}[z]=-\frac{1}{2l}\zeta(-z)$ corresponds to the energy
of the ground state (see below).

It is worth mentioning that eq. \eqref{chz} can be seen as a chiral
copy of the partition function of the (nonchiral) anisotropic free
boson found in \cite{Melnikov:2018fhb}, once the zero modes of the
latter are discarded.

\subsection{Microscopic counting of states and number theory}

The partition function $Z[\beta]$ can be alternatively obtained from
the trace over the Hilbert space of $\exp\left(-\beta\text{H}\right)$,
where H is the Hamiltonian operator, i.e.,
\begin{equation}
Z[\beta]=\text{Tr}\left[\exp(-\beta\text{H})\right]\,.\label{eq:Z- trace}
\end{equation}
The Hilbert space can be constructed out from the $\hat{u}(1)$ current
operators $\text{K}_{n}$ associated to \eqref{K}, whose algebra
is given by the quantum version of \eqref{U1}
\begin{equation}
[\text{K}_{n},\text{K}_{m}]=\pi n\delta_{n+m,0}\,.
\end{equation}
The occupation number states can then be defined as
\begin{equation}
|n_{1},n_{2},\cdots\rangle=\text{K}_{-1}^{n_{1}}\text{K}_{-2}^{n_{2}}\cdots|0\rangle\,,\label{eon}
\end{equation}
where $|0\rangle$ stands for the vacuum state that is annihilated
by $\text{K}_{n}$ with $n>0$.

In order to diagonalize the Hamiltonian operator that comes from \eqref{Ham},
we express it in the basis of normal ordered currents $\text{K}_{n}$,
which reads
\begin{equation}
\text{H}=\frac{1}{\pi l}\sum_{n=1}^{\infty}n^{z-1}\text{K}_{-n}\text{K}_{n}+\frac{1}{2l}\zeta(-z)\,.\label{eq:Ham-Operator}
\end{equation}
The $\hat{u}(1)$ descendants \eqref{eon} are then eigenstates of
the Hamiltonian \eqref{eq:Ham-Operator}, since
\begin{equation}
\text{H}|n_{1},n_{2},\cdots\rangle=\left(\sum_{k=1}^{\infty}n_{k}E_{k}-E_{0}[z]\right)|n_{1},n_{2},\cdots\rangle\,,
\end{equation}
with
\begin{equation}
E_{k}=k^{z}\,,
\end{equation}
and as aforementioned, $E_{0}[z]=-\frac{1}{2l}\zeta(-z)$ stands for
the ground state energy.

Therefore, the partition function \eqref{eq:Z- trace} can be explicitly
computed as
\begin{align}
Z & =\sum_{n_{1},n_{2},\cdots}\langle n_{1},n_{2},\cdots|\exp(-\beta\text{H})|n_{1},n_{2},\cdots\rangle\,,\\
 & =q^{\frac{1}{2}\zeta(-z)}\prod_{k=1}^{\infty}\sum_{n_{k}}\left(q^{E_{k}}\right)^{n_{k}}\,,\\
 & =q^{\frac{1}{2}\zeta(-z)}\prod_{k=1}^{\infty}\frac{1}{1-q^{k^{z}}}\,,\label{eq:Z-states}
\end{align}
in full agreement with the result from the path integral in \eqref{chz},
as expected. The equivalence of both ways of computing the partition
function is reassuring, since it means that the states have been well
identified and well counted.

The partition function \eqref{eq:Z-states} is defined in the canonical
ensemble, and it is useful to express it in terms of the density of
states at fixed energy
\begin{equation}
E=\sum_{i}E_{n_{i}}=\sum_{i}n_{i}^{z}=N\,.
\end{equation}
In order to count only along indistinguishable configurations, the
ordering $n_{1}\geq n_{2}\geq\cdots\geq0$ can be assumed. Hence,
following \cite{Melnikov:2018fhb}, the number of states with a fixed
energy $E$ is given by the ``power partitions'' $p_{z}(N)$, defined
as the number of partitions of an integer $N$ into $z$-th powers;
i.e., partitions of the form $N={\displaystyle \sum_{i}n_{i}^{z}=E}$.

Consequently, the sum over states can be rearranged, so that the partition
function can also be written as
\begin{align}
Z & =q^{\frac{1}{2}\zeta(-z)}\sum_{N}p_{z}(N)q^{N}\,.\label{eq:Z-N}
\end{align}
Indeed, the equivalence between the different ways of expressing the
partition function, as in \eqref{eq:Z-states} and \eqref{eq:Z-N},
holds due to an old and well-known identity in number theory, which
asserts that the sequence of power partitions, for a generic value
of $z$, has the following generating function (see e.g. \cite{hardy1918asymptotic,gafni2016power})
\begin{equation}
\sum_{N}p_{z}(N)q^{N}=\prod_{n=1}^{\infty}\frac{1}{1-q^{n^{z}}}\,.
\end{equation}

\subsection{Asymptotic growth of the number of states}

According to \eqref{eq:Z-N}, the entropy of a gas of non-interacting
anisotropic chiral bosons in the microcanonical ensemble is then exactly
given by
\begin{equation}
S=\log p_{z}(N)\,.\label{eq:S}
\end{equation}
Thus, for high temperatures, which corresponds to energies much greater
than the ground state, $E\gg E_{0}$, the entropy in \eqref{eq:S}
can be expressed in a closed form by virtue of the asymptotic growth
of the power partitions $p_{z}(N)$ with $N\gg1$.

The renowned formula for the asymptotic growth of the partitions $p_{1}(N)$
was found long ago by Hardy and Ramanujan in \cite{hardy1918asymptotic}
where they provided a thorough proof. Noteworthy, by the end of the
same work, they also conjectured a very precise formula for the asymptotic
growth of the power partitions, which reads
\begin{equation}
p_{z}(N)\approx(2\pi)^{-\frac{1}{2}(1+z)}\sqrt{\left(\frac{z}{1+z}\right)}k_{z}N^{\frac{1}{1+z}-\frac{3}{2}}\exp\left[(1+z)k_{z}N^{\frac{1}{1+z}}\right]\,,
\end{equation}
with
\begin{equation}
k_{z}=\left\{ \frac{1}{z}\Gamma(1+\frac{1}{z})\zeta(1+\frac{1}{z})\right\} ^{\frac{z}{1+z}}\,,
\end{equation}
whose accuracy was rigorously proved later in \cite{wright1934}.\footnote{Simplified proofs have been recently found in \cite{vaughan2015squares}
for $z=2$, and in \cite{gafni2016power,tenenbaum2019power} for $z\geq2$.
Extensions for non-integer values of $z$ have also been recently addressed in \cite{luca2016explicit,li2018r} (see also \cite{Gonzalez:2011nz,Melnikov:2018fhb}).}

Therefore, at high temperature, the entropy in \eqref{eq:S}, including
subleading corrections acquires the form
\begin{equation}
S=(1+z)k_{z}N^{\frac{1}{1+z}}-\frac{1}{2}\left(1+3z\right)\log N^{\frac{1}{1+z}}+\cdots
\end{equation}

\section{Ending remarks}

We have constructed the quantum field theory of a chiral free boson
with anisotropic scaling which, despite its simplicity, exhibits many
interesting features in common with those found in two-dimensional
conformal field theories. In this sense, it is worth highlighting
that, even though the anisotropic scaling manifestly breaks the relativistic
Lorentz symmetry, the standard conformal symmetry is still present,
but realized in a nonlocal way. Indeed, in the quantum theory, the
corresponding generators fulfill the Virasoro algebra
\begin{equation}
[\text{L}_{m},\text{L}_{n}]=(m-n)\text{L}_{m+n}+\frac{1}{12}m(m^{2}-1)\delta_{m+n,0}\,,
\end{equation}
independently of the dynamical exponent $z$, and so it agrees with
the standard result in the isotropic case.

It is also worth mentioning that the leading term of the entropy,
which is obtained from solid results in number theory, exactly agrees
with an extension of the Cardy formula that relies on modular invariance
for generic systems with anisotropic scaling \cite{Gonzalez:2011nz,Perez:2016vqo,Melnikov:2018fhb}.
This fact suggests that our partition function might correspond to
a modular form, at least for a suitable high/low temperature regime.

It would also be interesting to analyze the properties of the fermionic
version of the anisotropic chiral field, which fulfills the same field
equation, but it is described by the following action principle
\begin{equation}
S_{f}[\chi]=\int dtd\phi\,\chi(\partial_{t}\chi+\partial_{\phi}^{z}\chi)\,,
\end{equation}
being invariant under the anisotropic scaling in \eqref{eq:AnistropyZ},
provided that the Grassmann-valued field $\chi$ scales as $\chi\rightarrow\lambda^{-\frac{1}{2}}\chi$.

\acknowledgments We thank Marcela Cárdenas, Kristiansen Lara, Dmitry
Melnikov, Fábio Novaes, Alfredo P\'erez and Pablo Rodríguez for useful
discussions and comments. This research has been partially supported
by FONDECYT grants Nº 1161311, 1171162, 1181031, 1181496, 1181628,
3170772 and the grant CONICYT PCI/REDES 170052. The Centro de Estudios
Científicos (CECs) is funded by the Chilean Government through the
Centers of Excellence Base Financing Program of Conicyt.



\bibliographystyle{fullsort}

\end{document}